\documentclass[twocolumn,showpacs,preprintnumbers,amsmath,amssymb,superscriptaddress]{revtex4}

\usepackage{dcolumn}

\usepackage[american]{babel}
\usepackage{units}
\usepackage{textcomp} 
\usepackage{graphicx}
\usepackage{amsmath}
\usepackage{bm}

\begin{document}

\renewcommand{\figurename}{Fig.}

\title{Highly Selective Terahertz Bandpass Filters Based on Trapped Mode Excitation}

\author{Oliver Paul}
\email{paul@physik.uni-kl.de}
\affiliation{Department of Physics and Research Center OPTIMAS, University of Kaiserslautern, Germany
}%

\author{Ren\'{e} Beigang}
\affiliation{Department of Physics and Research Center OPTIMAS, University of Kaiserslautern, Germany
}%
\affiliation{Fraunhofer Institute for Physical Measurement Techniques IPM, Freiburg, Germany
}%

\author{Marco Rahm}
\affiliation{Department of Physics and Research Center OPTIMAS, University of Kaiserslautern, Germany
}%
\affiliation{Fraunhofer Institute for Physical Measurement Techniques IPM, Freiburg, Germany
}%


\begin{abstract}
We present two types of metamaterial-based spectral bandpass filters for the terahertz (THz)
frequency range. The metamaterials are specifically designed to operate for waves at normal incidence and
to be independent of the field polarization. The functional structures are embedded in films of
benzocyclobutene (BCB) resulting in large-area, free-standing and flexible membranes with low
intrinsic loss. The proposed filters are investigated by THz time-domain spectroscopy and show a pronounced
transmission peak with over \unit[80]{\%} amplitude transmission in the passband and a transmission
rejection down to the noise level in the stopbands. The measurements are supported by numerical
simulations which evidence that the high transmission response is related to the excitation of
trapped modes.
\end{abstract}



\maketitle


\section{Introduction}
In the last ten years, metamaterials have emerged to be powerful tools for the manipulation of
light on the subwavelength scale. The scientific interest has been primarily driven by the
possibility of creating materials with new electromagnetic properties not occurring in nature such
as e.\ g.\ negative index materials \cite{veselago1968,smith2000a}, invisibility cloaks and
transformation optics \cite{pendry2006,pendry2008}. However, metamaterials are not only of
scientific interest for their exotic properties. In the frequency range between 0.1 and 10 THz,
which is usually referred to as the THz gap, the lack of electromagnetic response of most natural
materials has substantially obstructed the development of functional components. For the THz
technology, metamaterials can play a crucial role for the conception of artificial optical
components since their electromagnetic properties can be exactly designed to match the
functionality of an envisioned optical component. In this context, several optical elements as e.\
g.\ wave plates \cite{averitt2009}, THz amplitude \cite{padilla2006b} and phase modulators
\cite{chen2009} and spatial modulators \cite{chan2009} have already been successfully demonstrated.
\begin{figure*}[]
   \begin{center}
    \includegraphics[width=6in]{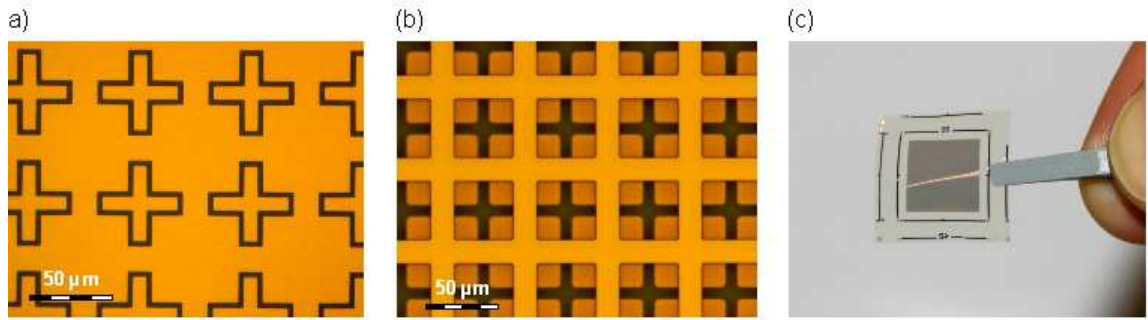}
    \end{center}
    \caption{Microscope pictures of (a) the cross-slot structure and (b) the two layers of wire-and-plate
    structure. (c) Resulting metamaterial membrane with a functional area of \unit[9$\times$9]{mm$^2$}.}
    \label{fig:structure}
\end{figure*}
Recently, the excitation of so-called trapped modes, i.e. modes that are weakly coupled to an
external electromagnetic field, has been observed in metamaterials
\cite{fedotov2007,zhang2008,Papasimakis2008,tassin2009,liu2009}. These modes show analogies to the
electromagnetically induced transparency (EIT) of atomic systems \cite{harris1990,liu2001} like a
sharp phase dispersion of the transmitted radiation and a narrow transmission band within a broad
stopband. Such properties open the possibility for the construction of very efficient and compact
metamaterial-based bandpass filters with a high selectivity.

In this paper, we present two types of spectral bandpass filters for the THz frequency range based
on the excitation of trapped modes. The corresponding resonances of the subwavelength elements have
been optimized to obtain a high transmission in the passband and an efficient suppression of
transmitted radiation in the stopbands. The two implemented metamaterial designs are a cross-slot
structure and a wire-and-plate structure (see Figs.~\ref{fig:structure}(a) and 1(b)). Such
structures have been originally introduced in the microwave regime \cite{munk2000,behdad2006}. They
operate at normal incidence and are independent of the polarization of the incident light. The
polarization insensitivity is a direct consequence of the 4-fold rotational symmetry of the
structure \cite{mackay1989}. In order to enhance the bandpass effect we employed a multilayer
technique to embed several functional layers in films of BCB. The BCB serves as a homogeneous
background matrix and enables us to fabricate large-area, free-standing and flexible metamaterial
membranes. This is especially important with regard to a practical integration of such metamaterial
components in THz systems since the beam diameter of THz radiation is usually in the order of
several millimeters. The designed and fabricated bandpass filters were experimentally characterized
by means of THz time-domain spectroscopy.

\section{Filter design und fabrication}
The cross-slot structure is set-up by an array of \unit[3]{$\mbox{\textmu}$m} wide cross-shaped
slots (Fig.~\ref{fig:structure}(a)). The enclosed crosses are formed by
\unit[46]{$\mbox{\textmu}$m} long and \unit[9]{$\mbox{\textmu}$m} wide cross bars. The cross bars
act as small electric dipoles that can be excited by an incident THz wave. The lattice constants of
the structure are \unit[68]{$\mbox{\textmu}$m} in the x- and y-direction and
\unit[40]{$\mbox{\textmu}$m} in the z-direction. In contrast, the wire-and-plate structure
(Fig.~\ref{fig:structure}(b)) is composed by two separated layers being
\unit[9.5]{$\mbox{\textmu}$m} apart from each other. The front layer consists of a two-dimensional
wire grid formed by \unit[17]{$\mbox{\textmu}$m} wide wires whereas the background layer is
represented by an array of square plates with a side length of \unit[50]{$\mbox{\textmu}$m}. The
lattice constants for this structure are \unit[60]{$\mbox{\textmu}$m} in the x- and y-direction and
\unit[35]{$\mbox{\textmu}$m} in the z-direction. Since the adjacent edges of each two plates act as
a capacitor whereas the facing strip of the wire grid acts as an inductor, the composite structure
forms an LC-resonant circuit that can be excited by a normally incident THz wave.

The fabrication of the metamaterial films was performed in a multilayer process with alternating
layers of BCB~3022-63 and copper on top of a silicon substrate. The BCB layers were fabricated by a
spin coating technique followed by a thermal curing process in a vacuum oven at
\unit[300]{$^\circ$C} for about \unit[5]{h}. The metal layers were patterned by standard
UV-lithography using an AZ~nLof~2035 photoresist, an EVG~620 mask aligner and an electron beam
evaporation of \unit[200]{nm} copper. For the plate-and-wire design a strict alignment of the
plates and wires layers within a unit cell is necessary to ensure the functionality of the
structure. For this purpose we used alignment marks providing an accuracy in the order of
\unit[1]{$\mbox{\textmu}$m}. A microscope image of one layer of unit cells of both designs is shown
in Figs.~\ref{fig:structure}(a) and~1(b), respectively. The films were then removed from the
silicon substrate in a \unit[30]{\%} solution of KOH. The resulting free-standing membranes are
\unit[17$\times$17]{mm$^2$} large, mechanically and chemically stable and quite flexible. A
photograph of the resulting membrane is presented in Fig.~\ref{fig:structure}(c).

We fabricated membranes with one layer of unit cells of the cross-slot structure and two layers of
unit cells of the wire-and-plate structure. However, as shown in \cite{paul2008}, the free-standing
membranes can be stacked on top of another to further increase the number of layers. Similar to the
double-cross structure reported in \cite{paul2008}, the structures used for the filter designs are
independent of the polarization and the coupling between the functional metal layers in neighboring
membranes can be neglected due to the thick BCB spacer. Hence, the alignment of individual
membranes is not crucial to the orientation or the relative position of the membranes and can be
performed under simple visual control.

\section{Results and discussion}
The transmission characteristics of the metamaterial filters was analyzed by standard THz
time-domain spectroscopy with a detectable frequency range of \unit[0.1 -- 2.5]{THz} and a
frequency resolution of \unit[9]{GHz}. The THz radiation was linearly polarized and was focused
under normal incidence on the sample surface to a spot size of \unit[1.5]{mm}. Finally, the
measured transmission spectra have been normalized by a reference spectrum without sample to obtain
the amplitude transmittance of the filters.

We analyzed one and two layers of unit cells of the cross-slot structure by measuring a single and
two stacked membranes, each fabricated with one layer of unit cells. For the wire-and-plate
structure we analyzed two and four layers of unit cells by using a single and two stacked membranes
where each membrane consisted of two layers of unit cells. The experimentally obtained spectral
transmission data were compared to numerical simulations which have been carried out by a
commercially available time-domain solver (CST Microwave Studio), where the BCB can be described by
a dielectric constant of $\epsilon = 2.67$ and a loss parameter of \linebreak $\tan \delta = 0.01$
\cite{paul2008}. However, we varied the permittivity of BCB in order to fit the numerical data to
the experimental results and obtained reasonable agreement by using $\epsilon = 2.45$ for the
cross-slot and $\epsilon = 1.85$ for the wire-and-plate design.

\begin{figure*}[t]
   \begin{center}
    \includegraphics[width=6in]{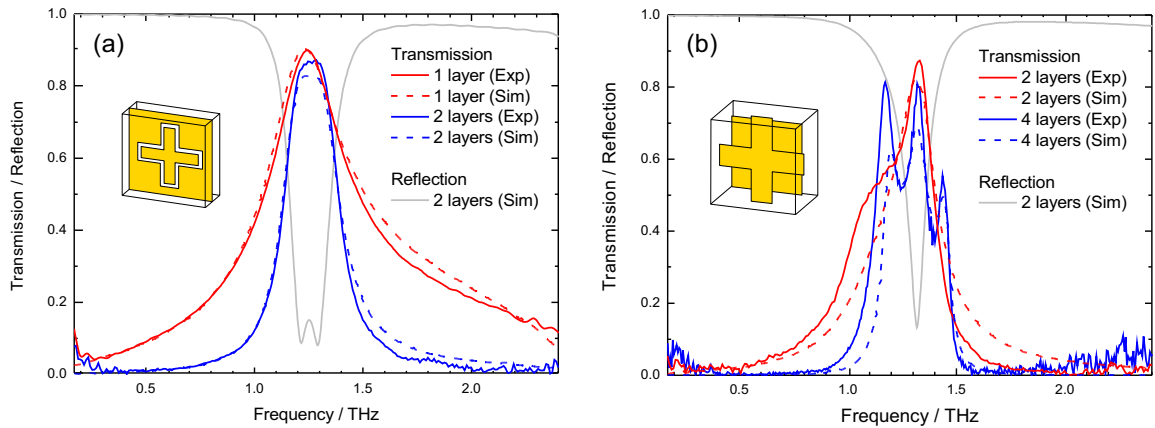}
    \end{center}
    \caption{Experimental (Exp) and numerical (Sim) amplitude transmission and reflection results for (a) the cross-slot and (b) the wire-and-plate structure for different numbers of layers of unit cells.}
    \label{fig:results}
\end{figure*}
\begin{figure*}[t]
   \begin{center}
    \includegraphics[width=6in]{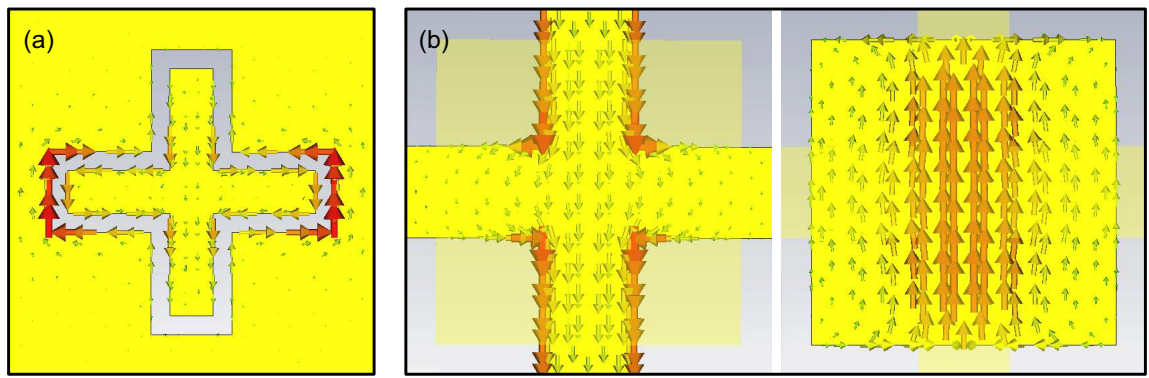}
    \end{center}
    \caption{Surface current distribution at the center frequency of the passband for (a) the cross-slot structure and (b) the front plane (left) and the backplane (right) of the wire-and-plate structure. The incident electric field is vertically polarized.}
    \label{fig:strom}
\end{figure*}

Figs.~\ref{fig:results}(a) and \ref{fig:results}(b) show the spectral amplitude transmission and
reflection of the cross-slot-structure and the wire-and-plate structure, respectively. The
experimental transmission results (colored solid lines) are in good agreement with the numerical
simulations (colored dashed lines). Both filter designs reveal a pronounced passband around
\unit[1.3]{THz}. As expected, the frequency selectivity of the bandpass filters increases with
increasing number of layers of unit cells. For two layers of the cross-slot structure and for four
layers of the wire-and-plate structure the FWHM bandwidth of the passband is $\Delta f =
\unit[0.3]{THz}$ in each case. Both filters offer a very high amplitude transmission over
\unit[80]{\%}, a fast roll-off and a very efficient blocking in the lower and upper rejection bands
down to the noise level where the incident radiation is almost completely reflected. Moreover, the
transmission response of the cross-slot structure is ripple-free, whereas the wire-and-plate filter
exhibits a faster roll-off.

The origin of these strong resonances can be attributed to so-called trapped modes
\cite{fedotov2007,zhang2008,Papasimakis2008,tassin2009,liu2009}, i.e. modes that are weakly coupled
to electromagnetic waves incident from free space. For such modes the radiation losses are very
small in comparison to the stored field energy which leads to an enhanced transmission at the
resonance frequency. The excitation of trapped modes is evidenced by the simulation of the surface
current distribution in the metamaterial structures presented in Fig.~\ref{fig:strom}. At the
resonance frequency, the induced currents are counter propagating at distinct sections of the
structure with almost similar magnitude. As a consequence, the resulting dipole moment and
therefore the dipolar coupling to external electromagnetic fields is strongly reduced which results
in a high transmission of electromagnetic radiation through the metamaterial structure at
frequencies near the resonance.

In particular, for the cross-slot structure the surface currents are counter propagating at the two
opposing edges of the slot (Fig.~\ref{fig:strom}(a)). This specific current distribution is related to the fact that the
outer metal frame is just the complementary of a cross structure which causes the driven currents
to oscillate with opposite phase. For the wire-and-plate structure, it's the currents in the
front layer (metal wire grid) and the background layer (square metal patches) that are in opposite
phase (Fig.~\ref{fig:strom}(b)). This can be explained by the different functions of the layers in the equivalent LC-resonant circuit. As
mentioned in Sec.~2, the plates act as capacitors, i.e. the phase of the driven currents is shifted
by $+\pi/2$ with respect to the external electric field, whereas the wires act as inductors which
causes a phase shift of $-\pi/2$. This implies that the excited currents in the two layers must be
opposite in phase.
\begin{figure*}[t]
   \begin{center}
    \includegraphics[width=5in]{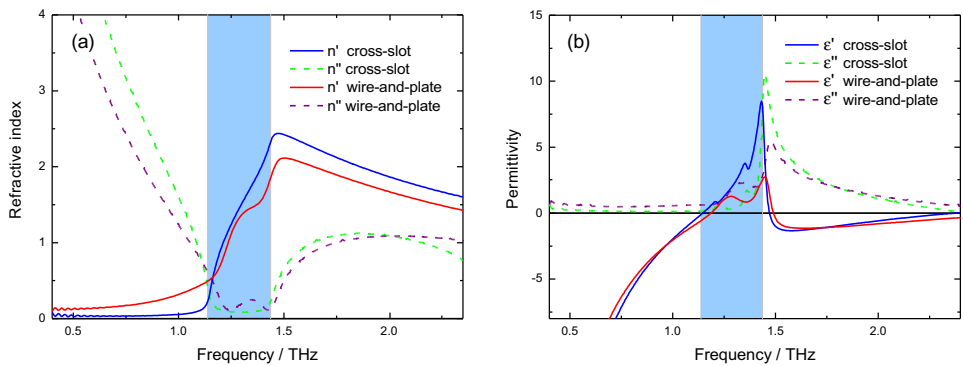}
    \end{center}
    \caption{Retrieved values of (a) the effective index of refraction and (b) the effective permittivity of the bandpass filters where ($\cdot$)' and ($\cdot$)'' denote the real and imaginary part, respectively. The spectral passband is shaded.}
    \label{fig:retrieval}
\end{figure*}
\section{Effective material parameters}
For a more quantitative characterization of the investigated metamaterial bandpass filters, we
further applied a retrieval algorithm \cite{chen2004} to calculate the effective values of the
refractive index~$n$ and the permittivity~$\epsilon$ from the simulated transmission and reflection
data. The retrieval was supported by additional computation of the phase advance of a propagating
plane wave across the material to ensure that the correct branch of the refractive index was
chosen.

The resulting effective refractive index and the permittivity are plotted in
Fig.~\ref{fig:retrieval} for both the cross-slot structure and the wire-and-plate structure.
Thereby, ($\cdot$)' and ($\cdot$)'' denote the real and imaginary part, respectively and the
spectral passband is indicated by a blue box. It can bee seen from Fig.~\ref{fig:retrieval}(b) that
the permittivity of both structures exhibits a characteristic narrow resonance, where $\epsilon'$
is only positive in the vicinity of the resonance frequency. As a consequence, only in this region
$n''$ is sufficiently small to allow high transmission leading to the observed passband. It should
be noted that both structures exhibit similar effective material parameters even though their
constituent elements widely differ in shape and geometry. This is due to the fact, that the
transmission response of both structures is related to the same origin: the excitation of trapped
modes in subwavelength elements. Moreover, since the quality factors of the excited resonances are
equal, the media that are composed by the two structures are equivalent in the framework of
effective medium theory. Another remarkable result is the rapid increase of $n'$ within the
passband as displayed in Fig.~\ref{fig:retrieval}(a). This strong frequency dispersion leads to an
increase of the group index which is given by $ n_g = \frac{c}{v_g}=c\frac{\partial k'}{\partial
\omega} = n'+\frac{\partial n'}{\partial \omega}\omega $ with the group velocity $v_g$ and the
dispersion relation $k'=n'\frac{\omega}{c}$. From the plotted curves for $n'$, the average group
index can be calculated in the passband to be $n_g=7.4$ for the wire-and-plate structure and even
$n_g=9.3$ for cross-slot structure. This means that a propagating pulse whose spectrum covers the
passband will be transmitted with a significant time delay. Although the group refractive index is
not as high as can be expected in the case of electromagnetically induced transparency
\cite{liu2001} or accordingly plasmon-induced transparency \cite{zhang2008} where a dark mode is
phase-coupled to a broadband dipole resonance, the calculations demonstrate the highly dispersive
character of trapped mode excitation.

\section{Conclusion}
In summary, we have presented two types of metamaterial bandpass filters in the THz frequency range.
The implemented metamaterials are based on a cross-slot and a wire-and-plate structure,
respectively. The filters are embedded in membranes of BCB allowing free-standing, flexible films
and are designed to operate at normal incidence and to be independent of the polarization of the
incident light. We have shown that the observed transmission response is related to the excitation
of trapped modes where the reduced coupling to the electromagnetic field leads to an enhanced
transmission at the resonance frequency.

The special characteristics of the presented filters is an outstanding high transmission over
\unit[80]{\%} in the passband and a fast roll-off down to the noise level in the stopbands. The
spectral bandwidth of the realized band-pass filters is \unit[0.3]{THz}. Such highly selective
filters can be used to remove unwanted transmitted signals in pre-defined frequency bands and have
potential applications in the field of THz diagnostics.
\\
\\
We thank Dr.~Christian Imhof from the Department of Electrical and Computer Engineering, University
of Kaiserslautern, for supportive comments and discussions, and the Nano+Bio Center at the
University of Kaiserslautern for their support in the sample fabrication.


\end{document}